\documentclass[preprint,sort&compress,3p,twocolumn]{elsarticle}
\usepackage{amssymb,amsbsy,amsmath,amsfonts,multirow}
\usepackage{yhmath}

\usepackage{slashed}
\usepackage{bm}
\usepackage{times}

\usepackage[dvipsnames,table,xcdraw]{xcolor}
\definecolor{lightyellow}{RGB}{255,250,205}

\usepackage{graphicx}
\graphicspath{{./}{./img/}{./fig/}{./image/}{./figure/}{./picture/}}
\usepackage{epstopdf}

\usepackage{tabularx}
\usepackage{longtable}

\usepackage{hyperref}
\hypersetup{
  colorlinks=true,        
  linkcolor=blue,         
  citecolor=cyan,         
  urlcolor=red,           
}

\begin{document}

\title{$K^*$ mesons with hidden charm arising from $KX(3872)$ and $KZ_c(3900)$ dynamics}

\author[X]{Xiu-Lei Ren}

\author[B]{Brenda B.~Malabarba}

\author[L1,L2]{Li-Sheng Geng}
\cortext[cor1]{lisheng.geng@buaa.edu.cn}

\author[K,L1]{K. P. Khemchandani}
\cortext[cor2]{kanchan.khemchandani@unifesp.br}


\author[B,L1]{A. Mart\'inez Torres}

\address[X]{Institut f\"{u}r Theoretische Physik II, Ruhr-Universit\"{a}t Bochum, D-44780 Bochum, Germany}
\address[B]{Instituto de F\'isica, Universidade de S\~ao Paulo, C.P. 66318, 05389-970 S\~ao 
Paulo, S\~ao Paulo, Brazil}
\address[L1]{School of Physics and Nuclear Energy Engineering \& Beijing Key Laboratory of Advanced Nuclear Materials and Physics, Beihang University, Beijing 100191, China}
\address[L2]{Beijing Advanced Innovation Center for Big Date-based Precision Medicine, Beihang University, Beijing100191, China}
\address[K]{Universidade Federal de S\~ao Paulo, C.P. 01302-907, S\~ao Paulo, Brazil}

\begin{abstract}
Inspired by the recent discovery of the pentaquark states $P_c(4450)$ and $P_c(4380)$, which can be viewed as excited nucleon states with hidden charm, we study
the three-body interaction of  a kaon and a pair of  $D\bar{D}^*$  in isospin 0 and 1.
We show that  the two body interactions stringently constrained  by the existence of the $D_{s0}^*(2317)$, $D^*_{s1}(2460)$, $X(3872)$, and $Z_c(3900)$,
which are widely believed to contain large $DK$, $D^* K$, and $D\bar{D}^*$ components, inevitably lead to
the existence of a heavy $K^*$ meson with hidden charm. Concrete coupled channel three-body calculations yield its mass
and width as  $(4307\pm 2)- i (9\pm 2)$ MeV with $I(J^P)=1/2(1^-)$. This state, if found experimentally, definitely cannot be
accommodated in a $q\bar{q}$ picture, and therefore presents a clear case of an exotic hadron.
\end{abstract}
\maketitle
\date{\today}

Understanding the nature of hadronic resonances/bound states is one of the most  challenging issues in the frontiers of hadron physics. 
In recent years,  experimental~\cite{Albrecht:2018xxd,Bhardwaj:2016lve,Bravina:2015cdt,Popov:2017dsy,Pappagallo:2016rye} 
and  theoretical~\cite{Guo:2017jvc,Chen:2016qju,Liu:2013waa,Olsen:2014qna,Hosaka:2016pey} efforts have been focusing
 on the nontraditional hadronic states, which cannot  be (easily) explained either as $q\bar{q}$ or $qqq$ states. One of the most recent claims on such kind of states is the existence of the $P_c(4380)$ and $P_c(4450)$ pentaquark states observed by the LHCb collaboration in the $J/\psi p$ invariant mass distribution of the $\Lambda^0_b\to J/\psi K^- p$ decay~\cite{Aaij:2015tga}.
Curiously the existence of such states of molecular $\bar D (\bar D^*)\Sigma_c/\Lambda_c$  nature was predicted prior to the experimental claim~\cite{Wu:2010jy}.

The possible existence of such non-conventional mesons and baryons dates back to the original quark model of Gell-Mann and Zweig, in which
 the existence of  multiquark states was already
anticipated~\cite{GellMann:1964nj, Zweig:1981pd}. In spite of such a long lapse of time, the recent intensified theoretical and experimental efforts  show clearly  that the topic is still controversial.

Regardless of all these efforts, there is still a vast unexplored energy region and systems in which states of non-conventional quark content could be found, especially at energies of 4$\sim$5 GeV. For instance, in the meson sector, heavy mesons of strangeness 0 with hidden charm, such as  $X(3872)$ or $Z_c(3900)$,  have been found, and they are widely regarded, particularly the $X(3872)$, as moleculelike states of $D \bar D^*$ in isospin 0 and 1, respectively (see, e.g., Refs.~\cite{Swanson:2003tb,Braaten:2003he,Gamermann:2006nm,Nieves:2012tt,Mehen:2011tp}). However, in the strange sector, there is surprisingly no experimental data available on heavy $K$ or $K^*$ meson states around this energy region, leaving the heavy strange physics experimentally unexplored. 
\begin{figure}[h!]
\begin{center}
\includegraphics[width=0.25\textwidth]{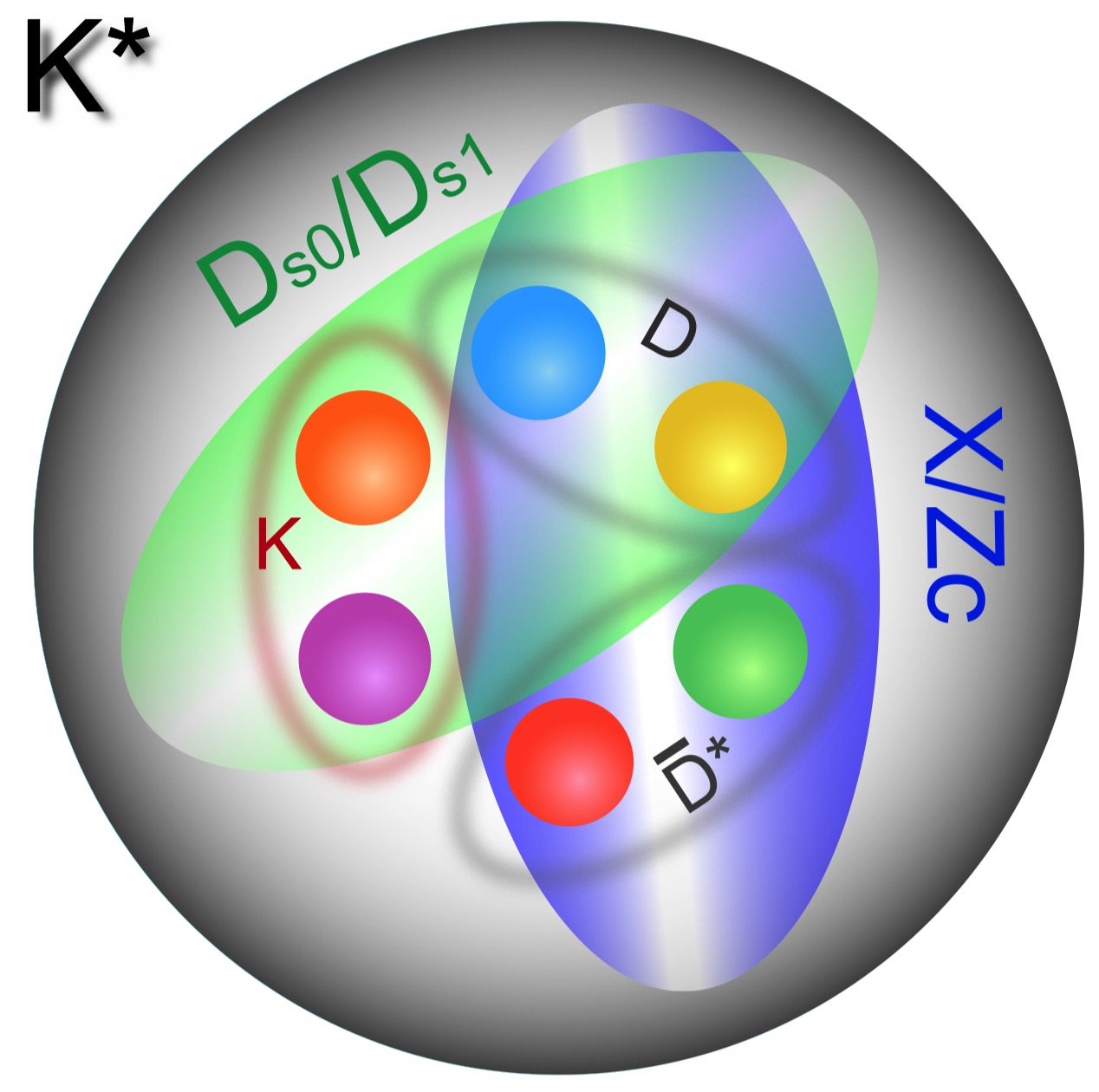}
\caption{Internal structure of the $K^*$ states found. The interaction $D\bar D^*$ forms the states $X(3872)$ in isospin 0 and $Z_c(3900)$ in isospin 1. When a $K$ is added to the system, the interaction between the $KD$ and $K\bar D^*$ systems is such that a bound state around 4300 MeV is formed whose internal structure corresponds to a $K$-$X$/$K$-$Z$ moleculelike state.}
\label{Fig:nature}
\end{center}
\end{figure}

In this letter, we explore the possibility of the existence of $K^*$ moleculelike states (bound states/resonances) with hidden charm in a three-meson system formed of a kaon and a pair of $D \bar D^*$, when the latter is organized either as $X(3872)$ or as $Z_c(3900)$. 
Different to other three-body studies,  the interactions of the two-body subsystems in the present case,
namely, the $DK$, $\bar{D}K$, $D^*K$, $\bar{D}^* K$,  and
$D\bar{D}^*$ are
stringently constrained by a large number of 
experimental as well as lattice QCD data. For instance,  it is a known fact that the $DK$,  $D^*K$,  and
$D\bar{D}^*$ interactions are  attractive such that  the   $D^*_{s0}(2317)$, $D^*_{s1}(2460)$,  $X(3872)$, and $Z_c(3900)$
can be understood as molecular states of the respective pair of hadrons~\cite{Gamermann:2006nm,Barnes:2003dj,Gamermann:2007fi,Guo:2006fu,vanBeveren:2003kd,Torres:2014vna,Bali:2017pdv,Albaladejo:2018mhb}.  In addition, studies in both
lattice QCD as well as chiral perturbation theory show that the $\bar D K $ and $\bar D^* K$ interactions in $I=0$ are moderately attractive while in $I=1$ are slightly repulsive~\cite{Liu:2012zya,Altenbuchinger:2013vwa}. 
Given such information, the existence of $K (D\bar{D}^*)$ bound states, depicted in Fig.~1,  depends on the possibility of the attractive pair interactions dominating over the repulsive ones. It is the purpose of the present work to confirm such a scenario by exploring the formation of resonances/bound states within
a concrete coupled-channel three body calculation. 

Furthermore, it is interesting to note the similarity of our system with one of the most studied three-body system, the $\bar{K}NN$ system (see, e.g., Refs.~\cite{Bayar:2012rk,Agakishiev:2014dha} and references
cited therein). 

In the present work, the $KD\bar{D}^*$ systems are studied  using the so-called fixed-center approximation (FCA)  to solve the Faddeev equations, where one of the two-body subsystem is considered as a scattering center, whose properties do not get altered during the scattering. Such a formalism is especially relevant to systems where two of the three hadrons form a bound state, while the third hadron is a light hadron as compared to the mass of the bound system. Indeed, FCA  has been successfully employed in describing the $\bar{K}d$ interaction at low energies~\cite{Chand:1962ec,Deloff:1999gc,Kamalov:2000iy}. Further, a comparison of the results obtained by solving the Faddeev equations, considering intermediate excitations of the bound system, and within FCA is done in Ref.~\cite{Toker:1981zh},  which shows  that the two results are very similar, implying that the FCA is a good approximation in such cases. More recently, the FCA to the Faddeev equations has been used to study the formation of three-hadron resonances in several system, such as $\phi K\bar{K}$, systems of one pseudoscalar and two vector mesons, $\eta K\bar{K}$ and $\eta' K\bar{K}$, $\pi \bar{K} K^*$, $\rho K\bar{K}$, $\rho D\bar{D}$,  $\rho D^* \bar{D}^*$, $\rho B^* B^*$, the $DKK$ and $DK\bar{K}$ systems, the $BDD$ and $BD\bar{D}$ systems (see, for example, Refs.~\cite{Dias:2017miz, Sekihara:2016vyd} and references therein).

Though eventually we will study the $KX(3872)$ and $KZ_c(3900)$ configurations of the three-body system in a coupled channel approach, we start by discussing the $KX(3872)$ scattering in detail. For this, we consider $X$ as a $(D \bar D^*)_{I=0}$ system. In such a picture, we can treat  $D\bar{D}^*$  as a cluster in the $KX(3872)$ scattering, which does not get perturbed by the low energy scattering of $K$  off the cluster, and apply the FCA to get the three-body scattering matrix $T$.  
\begin{figure}[ht]
\begin{center}
\includegraphics[width=0.48\textwidth]{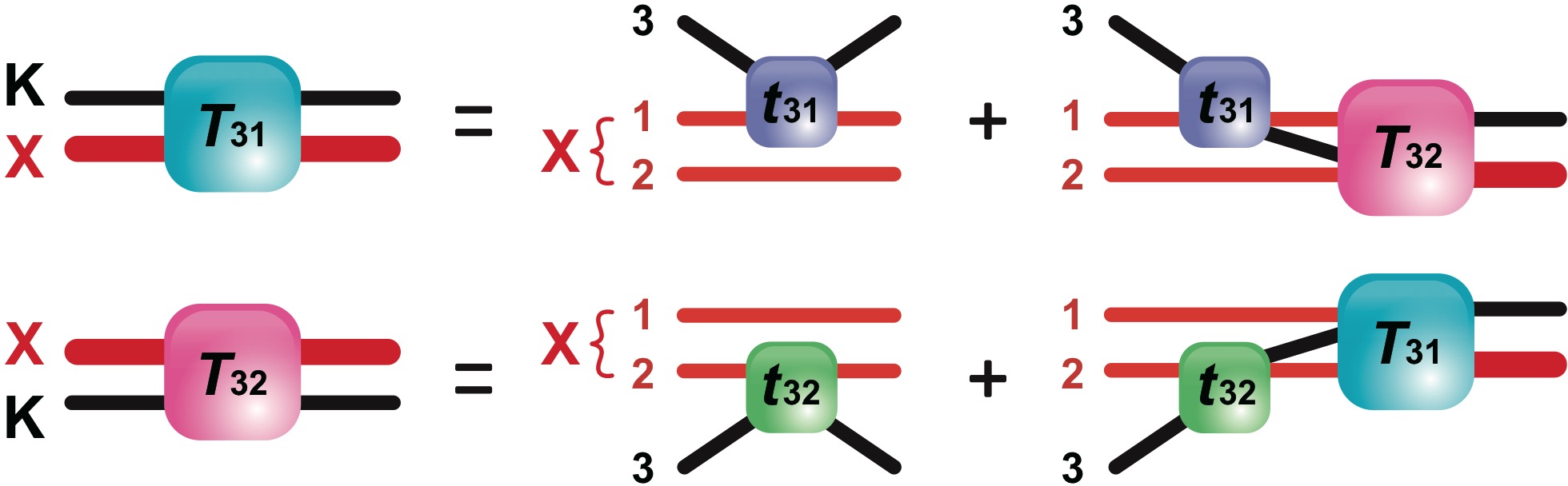}
\caption{Diagrams showing the scattering of the particle labeled ``3'' ($K$) on a cluster ($X$) made of particles 1 ($D$) and 2 ($\bar D^*$). }
\label{Fig:FCA}
\end{center}
\end{figure}
Within the FCA, $T$ can be written as a sum of two of the standard Faddeev partitions,  
\begin{align}
T=T_{31}+T_{32},\label{Tsumkx}
\end{align}
where $T_{31}$ and $T_{32}$ represent the sum of the infinite series of diagrams with the particle ``3" (kaon in the present case) first scattering off the particle ``1" ($D$) and ``2" ($\bar D^*$), respectively.  The sum of the two series are illustrated as diagrams in Fig.~\ref{Fig:FCA}, 
which can be expressed, mathematically, as a set of two coupled equations, 
\begin{eqnarray}\label{tpart}
  T_{31} &=& t_{31} + t_{31} G_0 T_{32},\\\nonumber
  T_{32} &=& t_{32} + t_{32} G_0 T_{31}.
 \end{eqnarray}
 
In Eq.~(\ref{tpart}), $t_{31}$, $t_{32}$  correspond to a weighted sum of the isospin 0 and 1 s-wave amplitudes of the $K D$ and $K\bar D^*$ systems. 
The weights related to the total isospin $I = 1/2$ configuration of the three-body system are summarized in Table~\ref{input} for the different transitions considered in this Letter. To obtain these weights we have to evaluate, for example, for $t_{31}$, 
\begin{align}
 \langle K C_b; I, I_z  |t_{31}|KC_a; I, I_z\rangle,   
 \end{align}
where $C_i$ is the $i$-th cluster, and  $I^i$, $I^i_z$ represent the isospin and its third component.

In this way, for example,  considering $D\bar D^*$ in isospin 0
\begin{align}
&|KX;1/2,1/2\rangle=|K^+;1/2,1/2\rangle\otimes |D\bar D^*;0,0\rangle\nonumber\\
&=\left|I^K_z=\frac{1}{2}\right\rangle\otimes\Bigg[\left|I^D_z=\frac{1}{2},I^{\bar D^*}_z=-\frac{1}{2}\right\rangle\nonumber\\
&\quad-\left|I^D_z=-\frac{1}{2},I^{\bar D^*}_z=\frac{1}{2}\right\rangle\Bigg],\label{KX1h}
\end{align}
and if we want to evaluate 
\begin{align}
\langle KX;1/2,1/2|t_{31}|KX;1/2,1/2\rangle,\label{KXtran}
\end{align}
we need to write the state in Eq.~(\ref{KX1h}) in terms of the total isospin of particles
3 and 1 (in this case, $K$ and $D$). Proceeding further, we obtain
\begin{align}
&|KX;1/2,1/2\rangle\nonumber\\
&=\frac{1}{2}\Bigg[\left|I^{KD}=1,I^{KD}_z=1\right\rangle\otimes\left|I^{\bar D^*}_z=-\frac{1}{2}\right\rangle\nonumber\\
&-\frac{1}{\sqrt{2}}\Big(\left|I^{KD}=1,I^{KD}_z=0\right\rangle\nonumber\\
&\quad+\left |I^{KD}=0,I^{KD}_z=0|\right\rangle\Big)\otimes\left|I^{\bar D^*}_z=\frac{1}{2}\right\rangle\Bigg].\label{idem}
\end{align}
Using Eq.~(\ref{idem}), Eq.~(\ref{KXtran}) gives
\begin{align}
&\langle KX;1/2,1/2|t_{31}|KX;1/2,1/2\rangle\nonumber\\
&=\frac{1}{4}(3t^{I=1}_{KD}+t^{I=0}_{KD})\nonumber\\
&\equiv \pmb{\omega}_{31}^{KX\to KX}\cdot \pmb{t}_{31},
\end{align}
where,
\begin{align}
&\pmb{\omega}_{31}^{KX\to KX}\equiv\frac{1}{4}\left(\begin{array}{cc}3&1\end{array}\right),\nonumber\\
&\pmb{t}_{31}\equiv\left(\begin{array}{c}t^{I=1}_{KD}\\\\t^{I=0}_{KD}\end{array}\right).\label{t31v}
\end{align}
Similarly, 
\begin{align}
&\langle KX;1/2,1/2|t^{32}|KX;1/2,1/2\rangle\nonumber\\
&\quad\equiv \pmb{\omega}_{32}^{KX\to KX}\cdot \pmb{t}_{32}\nonumber\\
\end{align}
with
\begin{align}
\pmb{\omega}&_{32}^{KX\to KX}=\pmb{\omega}_{31}^{KX\to KX},\nonumber\\
\pmb{t}_{32}&\equiv\left(\begin{array}{c}t^{I=1}_{K\bar D^*}\\\\t^{I=0}_{K\bar D^*}\end{array}\right).\label{t32v}
\end{align}
In general, we can write
\begin{align}
\langle K C_b; I, I_z  |t_{31}|KC_a; I, I_z\rangle&=\pmb{\omega}^{KC_a\to KC_b}_{31}\cdot \pmb{t}_{31},\nonumber\\
\langle K C_b; I, I_z  |t_{32}|KC_a; I, I_z\rangle &=\pmb{\omega}^{KC_a\to KC_b}_{32}\cdot \pmb{t}_{32}.\label{ww}
\end{align}
In Table~\ref{input}, we give the weight vectors $\pmb{\omega}_{31}$ and $\pmb{\omega}_{32}$  for the different transitions studied in this Letter.
The $KD$, $K\bar D^*$ $t$-matrices appearing in Eqs.~(\ref{t31v}) and (\ref{t32v}) are obtained by solving the Bethe-Salpeter equation in a coupled channel approach, using a kernel obtained from a Lagrangian based on heavy-quark spin symmetry.  As mentioned earlier, these coupled channel interactions generate the resonances $D^*_{s0}(2317)$ and $D^*_{s1}(2460)$. A normalization factor $\sqrt{M_aM_b}/m_{D(D^*)}$ is included, with $M_a$, $M_b$ being the masses related to the clusters in the initial and final states, respectively, in the definition of $t_{31}$ and $t_{32}$. The origin of this factor, as explained in Refs.~\cite{Roca:2010tf, MartinezTorres:2010ax}, lies in relating the $S$-matrix of the three-body system with the scattering of one particle on a cluster of the remaining two: considering a box of volume $\mathcal{V}$ in which the plane wave states are normalized to unity, the $S$-matrix related to the process $K+(c_1+c_2)\to K+(c_1+c_2)$, with $c_1$ and $c_2$ being the particles forming the cluster $C$, is given by
\begin{align}
S_{K+(c_1+c_2)}&=\delta_{\pmb{p}_K,~\pmb{p}^\prime_K}\delta_{\pmb{p}_{c_1},~\pmb{p}^\prime_{c_1}}\delta_{\pmb{p}_{c_2},~\pmb{p}^\prime_{c_2}}\nonumber\\
&\quad-i\frac{(2\pi)^4}{\mathcal{V}}\delta^{(4)}(P-P^\prime)\left(\prod_{i=1}^{3}\sqrt{\frac{N_i}{2E_i}}\right)\nonumber\\
&\quad\times\left(\prod_{j=1}^{3}\sqrt{\frac{N_j}{2E^\prime_j}}\right)T_{K+(c_1+c_2)},
\end{align}
where $P=p_K+p_{c_1}+p_{c_2}$ ($P^\prime=p^\prime_K+p^\prime_{c_1}+p^\prime_{c_2}$) is the initial (final) four-momentum, the index $i$ ($j$) represents the particles in the initial (final) state, $N_i$ is a normalization factor (1 for mesons and $2M_i$ for baryons of mass $M_i$), $E_i$ and $\pmb{p}_i$ ($E^\prime_i$ and $\pmb{p}^\prime_i$) correspond to the energy and 3-momentum, respectively, of the particle $i$ in the initial (final) state, and $T_{K+(c_1+c_2)}$ is the $T$-matrix related to the process. However, for a process $K+C\to K+C$, the $S$-matrix associated with it would be related to the $T$-matrix as
\begin{align}
S_{K+C}&=\delta_{\pmb{p}_K,~\pmb{p}^\prime_K}\delta_{\pmb{p}_{C},~\pmb{p}^\prime_{C}}-i\frac{(2\pi)^4}{\mathcal{V}}\delta^{(4)}(P-P^\prime)\nonumber\\
&\quad\times\left(\prod_{i=1}^{2}\sqrt{\frac{N_i}{2E_i}}\right)\left(\prod_{j=1}^{2}\sqrt{\frac{N_j}{2E^\prime_j}}\right)T_{K+C}.
\end{align}
This issue related to the normalization of the fields, as shown in Refs.~\cite{Roca:2010tf, MartinezTorres:2010ax}, can be solved by substituting $t_{31}$, $t_{32}$ and $G_0$ appearing in Eq.~(\ref{tpart}) by
\begin{align}
t_{31}&\to \sqrt{\frac{2E_C}{N_C}} \sqrt{\frac{2E^\prime_C}{N_C}} \sqrt{\frac{N_{c_1}}{2E_{c_1}}}\sqrt{\frac{N_{c_1}}{2E^\prime_{c_1}}}t_{31},\nonumber\\
t_{32}&\to \sqrt{\frac{2E_C}{N_C}} \sqrt{\frac{2E^\prime_C}{N_C}} \sqrt{\frac{N_{c_2}}{2E_{c_2}}}\sqrt{\frac{N_{c_2}}{2E^\prime_{c_2}}}t_{32},\nonumber\\
G_0&\to\sqrt{\frac{N_C}{2E_C}}\sqrt{\frac{N_C}{2E^\prime_C}}G_0.\label{norm}
\end{align}
Since all the particles involved are mesons, and using in Eq.~(\ref{norm}) the non-relativistic approximation of the energy $E_i~\sim M_i$ for heavy mesons of mass $M_i$, we have
\begin{align}
t_{31}&\to \frac{M_C}{M_{c_1}}t_{31},\quad t_{32}\to \frac{M_C}{M_{c_2}}t_{32},\nonumber\\
G_0&\to \frac{1}{2M_C}G_0.\label{tGnorm}
\end{align}
The normalization factors in $t_{31}$ and $t_{32}$ can be reabsorbed in the weight vectors $\pmb{\omega}_{31}$ and $\pmb{\omega}_{32}$ of Eq.~(\ref{ww}),
\begin{align}
\pmb{\omega}^{KC_a\to KC_b}_{31}\to \frac{\sqrt{M_aM_b}}{m_{D}}\pmb{\omega}^{KC_a\to KC_b}_{31},\nonumber\\
\pmb{\omega}^{KC_a\to KC_b}_{32}\to \frac{\sqrt{M_a M_b}}{m_{\bar D^*}}\pmb{\omega}^{KC_a\to KC_b}_{32}.\label{wnorm}
\end{align}

\begin{table}
\renewcommand*{\arraystretch}{2}
\caption{Weight vectors $\pmb{\omega}_{31 (32)}$ appearing in Eq.~(\ref{ww}) for total isospin $1/2$ of the three-body system and for the different transitions studied in this Letter. The first (second) element of the vectors $\pmb{\omega}_{31}$ and $\pmb{\omega}_{32}$ represents, respectively, the weight of the isospin one (zero) of the $t_{KD}$ and of the $t_{K\bar D^*}$ amplitudes (see Eqs.~(\ref{t31v}) and ~(\ref{t32v})). For the non-diagonal transitions, $\pmb{\omega}_{32}$ has to be multiplied by the factor $-1$.}\label{input}
\vspace{0.2cm}
\begin{tabular}{c|c|c}
\hline\hline
&$KX$&$KZ$\\
\hline
$KX$&$\frac{1}{4}\left(\begin{array}{cc}3&1\end{array}\right)$&$\frac{\sqrt{3}}{4}\left(\begin{array}{cc}1&- 1\end{array}\right)$
\\\hline
$KZ$& $\frac{\sqrt{3}}{4}\left(\begin{array}{cc}1&- 1\end{array}\right)$ &~$\frac{1}{4}\left(\begin{array}{cc}1&3\end{array}\right)$\\
\hline\hline
\end{tabular}
\end{table}

The loop function $G_0$, in Eq.~(\ref{tpart}), represents the Green's function of the $K$ meson propagating in the  $(D\bar{D}^*)_{I} $  cluster, and can be expressed as (including the normalization factor of Eq.~(\ref{tGnorm}))
\begin{align}\label{gzero}
& \langle K C_a; I, I_z | G_0 |K C_a; I, I_z \rangle \nonumber\\
 &\quad= \frac{1}{2M_a} \int \frac{d^3 \bm{q}}{(2\pi)^3}  \frac{F_a(\bm{q})}{{q^0}^2-\bm{q}^2 - m_{K}^2 + i\epsilon},
\end{align}
where $m_K$ represents the mass of the kaon and $q_0$ is the on-shell energy of the kaon in the center-of-mass frame of the kaon and the cluster:
\begin{equation}
  q^0 = \frac{s+m_{K}^2-M_a^2}{2\sqrt{s}},
\end{equation}
with $\sqrt{s}$ being the total energy. Note, that a form factor $F_a(\bm{q})$ is introduced in Eq.~(\ref{gzero}), which is related to the wave function of the cluster  in terms of its internal $D\bar D^*$ structure. We calculate this form factor following Refs.~\cite{MartinezTorres:2010ax,Roca:2010tf}  as
\begin{align}
  F_a(\bm{q})&= \frac{1}{\mathcal{N}} \int\limits_{|\bm{p}|,~ |\bm{p}-\bm{q}|<\Lambda} d^3 \bm{p} f_a(\pmb{p})f_a(\pmb{p-q}),\\
  f_a(\bm{p})&= \frac{1}{\omega_{D}(\bm{p})\omega_{\bar{D}^*}(\bm{p})}\frac{1}{M_a-\omega_{D}(\bm{p})-\omega_{\bar{D}^*}(\bm{p})}, \label{Eq:FF}
\end{align}
where  $\mathcal{N}=F_a(\bm{q}=\bm{0})$ is the normalization factor, and $\omega_{D}(\bm{p})=\sqrt{m_{D}^2+\bm{p}^2}$, $\omega_{\bar{D}^*}(\bm{p})=\sqrt{m_{\bar{D}^*}^2+\bm{p}^2}$.  The upper integration limit $\Lambda$ is chosen to be the same as the cutoff used to regularize the loop $D\bar{D}^*$ to get the cluster ($X(3872)$ or $Z_c(3900)$). We take $\Lambda\sim700$ MeV from Refs.~\cite{Gamermann:2006nm,Aceti:2012cb,Aceti:2014uea} and vary it up to $750$ MeV to estimate the uncertainties involved in the results. 

Using Eqs.~(\ref{Tsumkx}) and (\ref{tpart}), the total amplitude $T$ can be written as $T=T_{31}+T_{32}$, with
\begin{align}
T_{31} &=\left[ 1- t_{31} G_0 t_{32} G_0 \right]^{-1} [t_{31} + t_{31} G_0 t_{32}],\nonumber\\
T_{32} &=\left[ 1 - t_{32} G_0 t_{31} G_0 \right]^{-1} [t_{32} + t_{32} G_0 t_{31}],\label{fullt}
\end{align}
and is calculated as a function of the three-body invariant mass, $\sqrt{s}$. For a given $\sqrt{s}$, the two-body amplitudes are obtained at the invariant masses $s_{31}$ and $s_{32}$ of the relevant subsystem~\cite{Xie:2010ig}.

\begin{figure}[t]
\begin{center}
\includegraphics[width=0.5\textwidth]{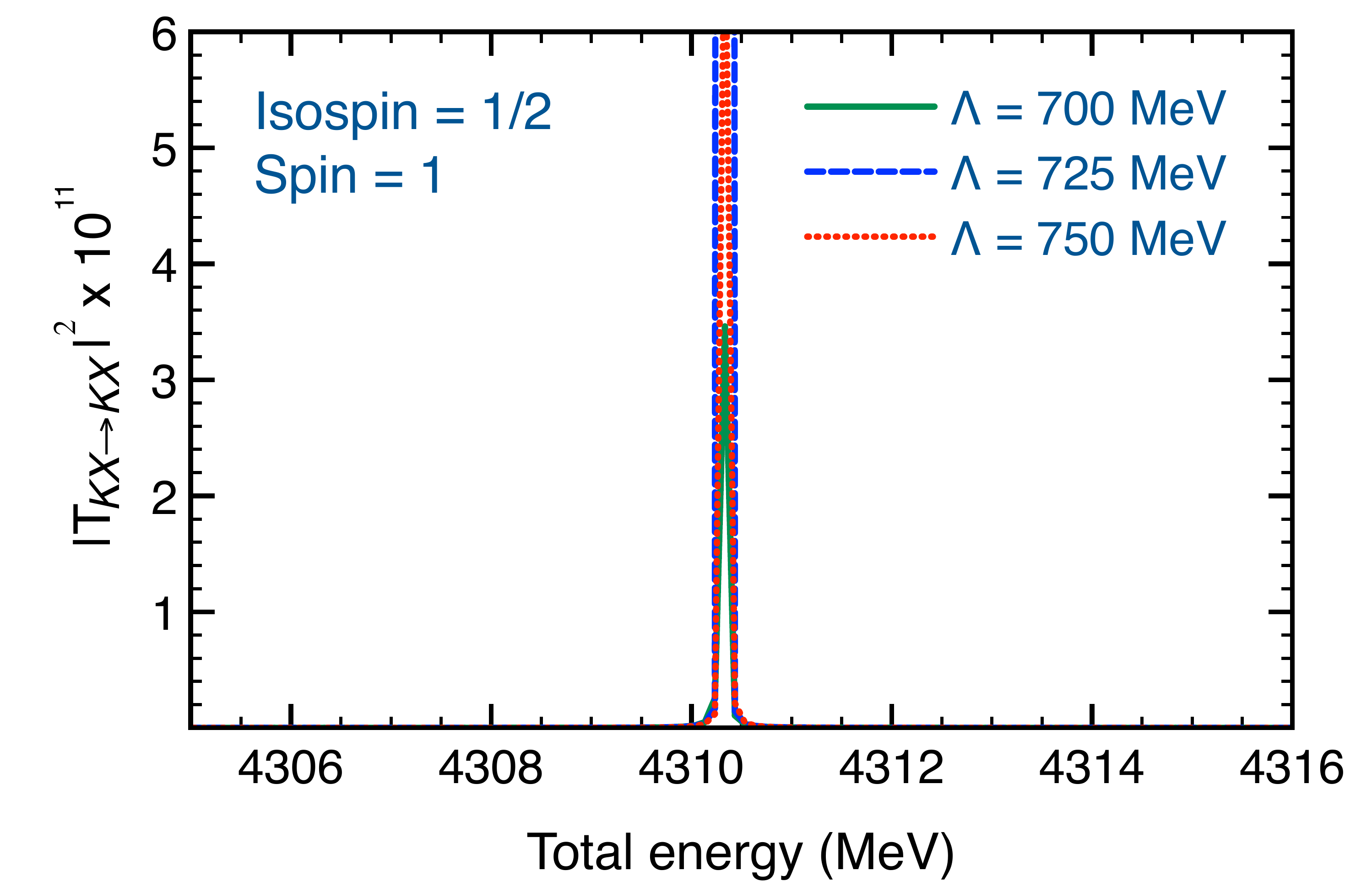}
\caption{Modulus squared of the $KX(3872)$ scattering amplitude. The solid, dashed, and dotted lines represent the results obtained with the cutoff $\Lambda=700$, $725$, and $750$ MeV, respectively.}
\label{Fig:KXamp}
\end{center}
\end{figure}

In Fig.~\ref{Fig:KXamp} we show the results found for the $T$-matrix of the $KX$ system for isospin $1/2$ and spin-parity $J^P = 1^-$. It can be seen from Fig.~\ref{Fig:KXamp} that a narrow peak appears around 4310 MeV, which almost does not vary with the cut-off.  

In the last years, the existence of several exotic companions of the $X(3872)$ has been claimed experimentally as well as theoretically (for reviews, see Refs.~\cite{Hosaka:2016pey,Olsen:2017bmm}). Particularly, $Z_c$ states, with isospin $1$, have been reported in the same energy region of the $X (3872)$, like the  $Z_c(3900)$ found by the BESIII~\cite{Ablikim:2015tbp}, or the $Z_c(3894)$ claimed by the Belle collaboration~\cite{Liu:2013dau} or the $Z_c(3886)$ reported by the CLEO collaboration~\cite{Xiao:2013iha}. At the present moment it is unclear, given the experimental uncertainties in the masses and widths, if all these experimental findings do, or do not, correspond to the manifestation of the same state. Such a discussion is beyond the scope of the present work, but it would be interesting to study under the same formalism as for $KX$ the existence of $K^*$ with hidden charm which could be interpreted as $KZ_c$ moleculelike states. Due to the present experimental uncertainty, we are using the name $Z$  to denote the isospin 1 partner of $X$ found in Ref.~\cite{Aceti:2014uea}, which has a mass around 3872 MeV and width around 30 MeV. In case of the scattering of $K$ with $Z$, to obtain reliable results, the width, $\Gamma$, of the $Z$ can play a relevant role. In our formalism such information can be introduced by replacing the mass $M$ of the cluster with  $M-i\Gamma/2$ in the expression of the form factor. Since $\Gamma_Z\sim 28$ MeV (compatible with the fit to the experimental data done in Ref.~\cite{Aceti:2014uea} and from the experimental data summarized in Ref.~\cite{Patrignani:2016xqp}) is not too large, and we are interested in studying the formation of states below the $KZ$ threshold, we can still rely on the FCA formalism to calculate the $KZ\to KZ$ amplitude. 

\begin{figure}[h]
\begin{center}
\includegraphics[width=0.47\textwidth]{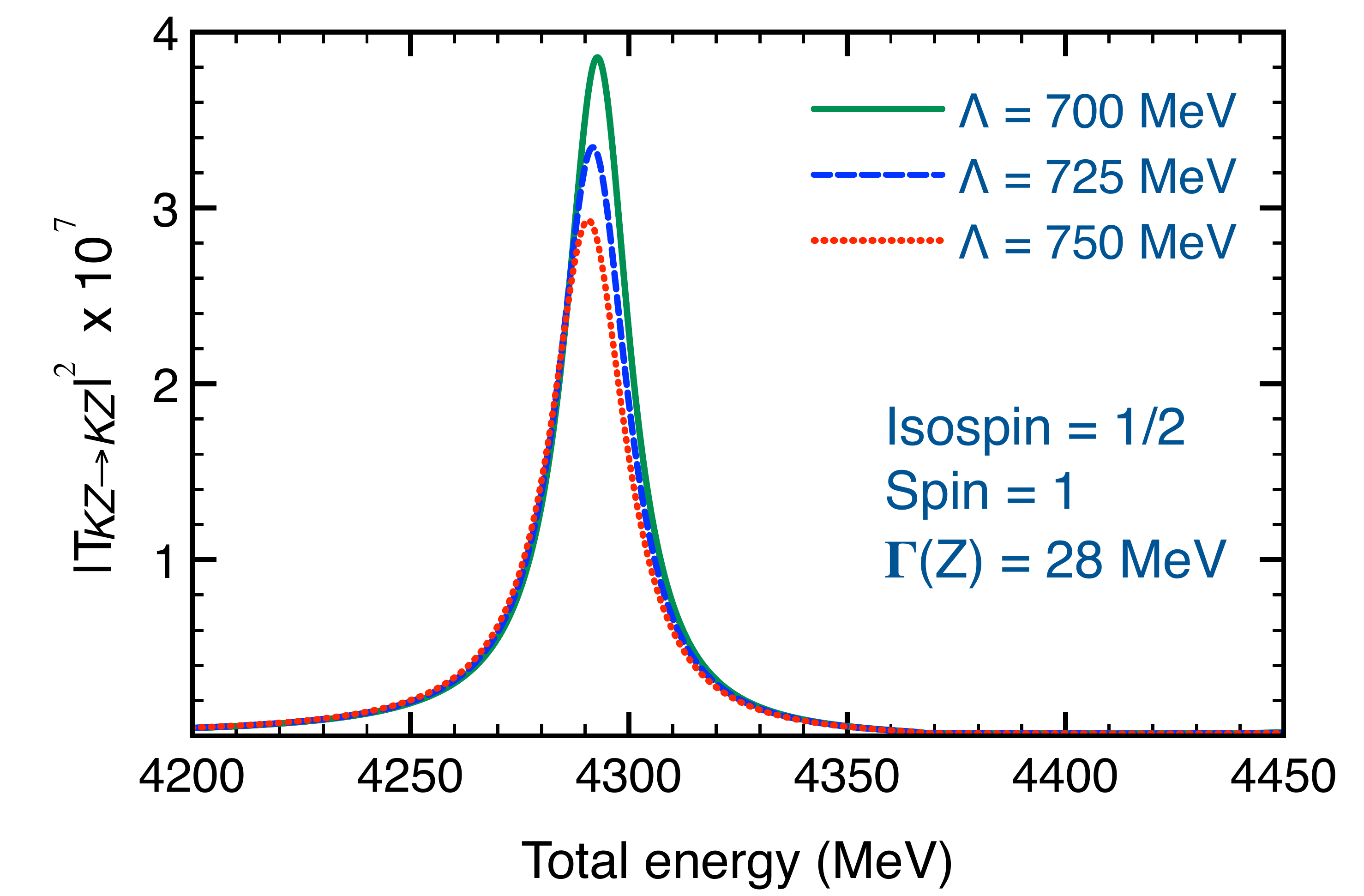}
\caption{Modulus squared of the $KZ$ scattering amplitude in $I=1/2$.}
\label{Fig:KZamp}
\end{center}
\end{figure}

In Fig.~\ref{Fig:KZamp} we show the modulus squared amplitude for $KZ$ scattering in isospin 1/2 (see Table~\ref{input} for the input two body $t$-matrices used in Eq.~(\ref{fullt})). A clear signal for the formation of a state around 4292 MeV and a width of 20 MeV is seen. If we neglect the width of the $Z$ state, a peak at $\sim$4300 MeV with a small width\footnote{\label{foot}The origin of this small width, even though the peak position lies below the $KZ$ and $K D \bar D^*$ thresholds, comes from the intermediate open channels, like, $\pi D_s \bar D^*$, which are implicitly considered in our formalism through the input $KD$ amplitude in isospin 1. This amplitude is obtained by solving the Bethe-Salpeter equation considering $KD$ and $\pi D_s$ as coupled channels. In fact, if the coupling to two-body open channels is switched off when getting the $KD$, $K\bar D^*$ amplitudes in isospin 1, we indeed find a zero width state in the $T$-matrix.}, $\sim$ 1 MeV, is observed.
In both cases, the mass of the state is about 70 MeV below the $KZ$ threshold (considering $Z$ as a stable particle). This energy region is well within the range of the reliability of the results obtained within the FCA.  The result obtained is also very stable with the cut-off $\Lambda$, as can be seen in Fig.~\ref{Fig:KZamp}. Thus we find an isospin 1/2, $J^P=1^-$, state with $M - i\Gamma/2 = 4292 - i10$~MeV in the $KZ$ scattering.  

The $KZ$ system can also have total isospin 3/2. If a state appears in this case, it would be associated with an exotic strange meson with isospin 3/2 and spin-parity $1^-$. We have studied this configuration of the $KZ$ system but find no state formed in it.  

Comparing the results of the $KX$ and $KZ$ systems in isospin 1/2, it can be concluded that both interactions result in formation of a state in the same energy region. However, the  $D \bar D^*$ system can reorganize itself in different isospin configurations during the scattering with kaon, while conserving the total isospin of the three-body system producing transitions between the configurations $KX$ and $KZ$, and due to the similar mass of $X$ and $Z$, the state found around 4300 MeV should have sizable internal $KX$ and $KZ$ structures. Such a possibility can be studied by treating $KX$ and $KZ$ as coupled channels, as done in Ref.~\cite{Xie:2010ig} for the state $N^*(1910)$, which can be considered as a molecular state with important $Nf_0(980)$ and $N a_0(980)$ components in its wave function. In such a case the $t_{31}$, $t_{32}$ and $G_0$ appearing in Eq.~(\ref{fullt}) are matrices in the coupled channel space\footnote{Note that Eq.~(\ref{fullt}), when written in a matrix form in terms of the matrices given in Eq.~(\ref{tGmat}), represent a more compact notation for writting the set of coupled equations (3)-(12) given in Ref.~\cite{Xie:2010ig}.}:
\begin{align}
t_{31}&=\left[\begin{array}{cc} (t_{31})_{11}&(t_{31})_{12}\\(t_{31})_{21}&(t_{31})_{22}\end{array}\right],\nonumber\\
t_{32}&=\left[\begin{array}{cc} (t_{32})_{11}&(t_{32})_{12}\\(t_{32})_{21}&(t_{32})_{22}\end{array}\right],\nonumber\\
G_{0}&=\left[\begin{array}{cc} (G_0)_{11}&0\\0&(G_0)_{22}\end{array}\right],\label{tGmat}
\end{align}
where the element $(11)$ represents $KX\to KX$, the element $(12)$ $KX\to KZ$, and so far so on. As done earlier, these $t$-matrices can be written  as
\begin{align}
(t_{31})_{ij}=\pmb{\omega}^{i\to j}_{31}\cdot \pmb{t}_{31},\nonumber\\
(t_{32})_{ij}=\pmb{\omega}^{i\to j}_{32}\cdot \pmb{t}_{32}.
\end{align}
The weight vectors $\pmb{\omega}_{31}$ and $\pmb{\omega}_{32}$ related to these processes can be found in Table~\ref{input} (without including the normalization factor discussed in Eq.~(\ref{wnorm})). Using this Table, for example, the element (12) of the $t_{31}$ and $t_{32}$ matrices is given by
\begin{align}
&(t_{31})_{12}=\pmb{\omega}^{KX\to KZ}_{31}\cdot \pmb{t}_{31}\nonumber\\
&\quad=\frac{\sqrt{M_XM_Z}}{m_D}\frac{\sqrt{3}}{4}(t^{I=1}_{KD}-t^{I=0}_{KD}),\nonumber\\
&(t_{32})_{12}=\pmb{\omega}^{KX\to KZ}_{32}\cdot \pmb{t}_{32}\nonumber\\
&\quad=-\frac{\sqrt{M_XM_Z}}{m_{\bar D^*}}\frac{\sqrt{3}}{4}(t^{I=1}_{K\bar D^*}-t^{I=0}_{K\bar D^*}).\nonumber
\end{align}

In such an approach, the Faddeev partitions $T_{31}$ and $T_{32}$ appearing in Eq.~(\ref{fullt}) are also matrices in the coupled channel space, such that Eq.~(\ref{fullt}) becomes a matrix equation and the $T$-matrix for the system is given by
\begin{align}
T=T_{31}+T_{32}=\left[\begin{array}{cc}T_{11}&T_{12}\\T_{21}&T_{22}\end{array}\right],
\end{align}
with $T_{11}$ ($T_{22}$) being the $T$-matrix for the $KX\to KX$ ($KZ\to KZ$) transition considering the coupled channel effect.

\begin{figure}[h!]
\begin{center}
\includegraphics[width=0.47\textwidth]{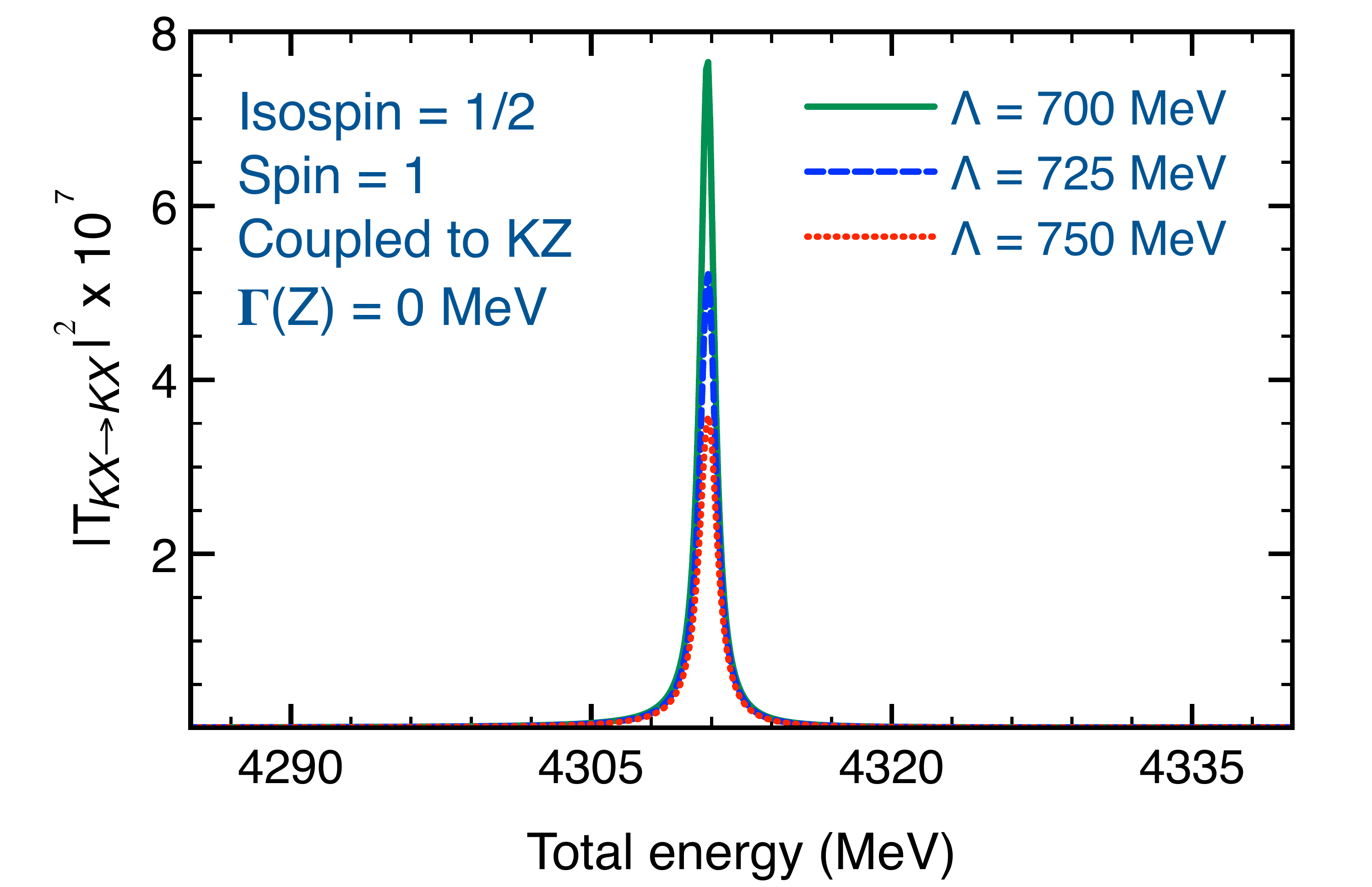}  \includegraphics[width=0.47\textwidth]{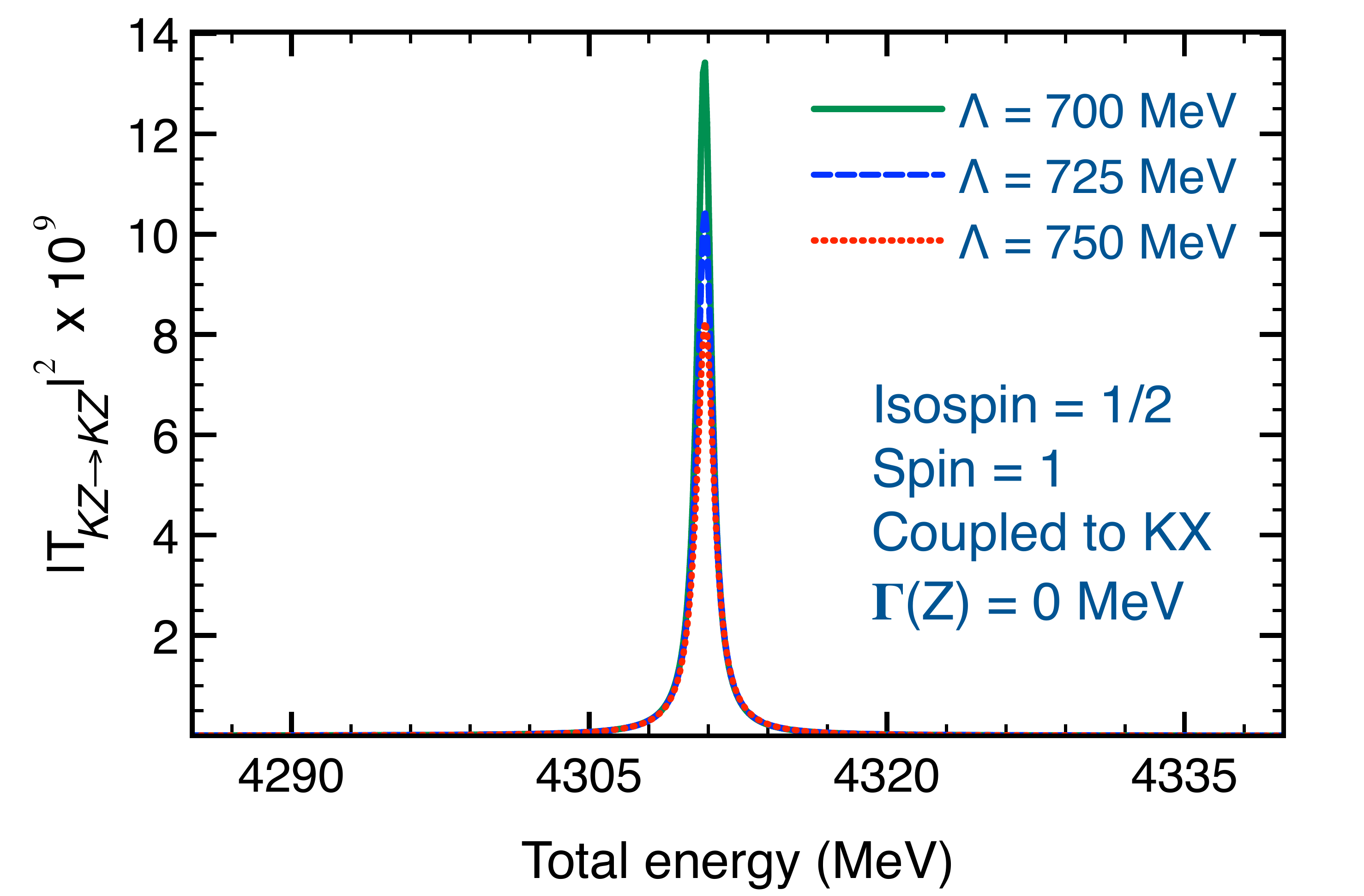}
\caption{Modulus squared of the $KX$ and $KZ$ scattering amplitudes in $I=1/2$. These results have been obtained by solving scattering equations while treating $KX$ and $KZ$ as coupled channels and considering $\Gamma(Z)=0$ MeV.}
\label{Fig:KXKZno}
\end{center}
\end{figure}

\begin{figure}[h!]
\begin{center}
\includegraphics[width=0.47\textwidth]{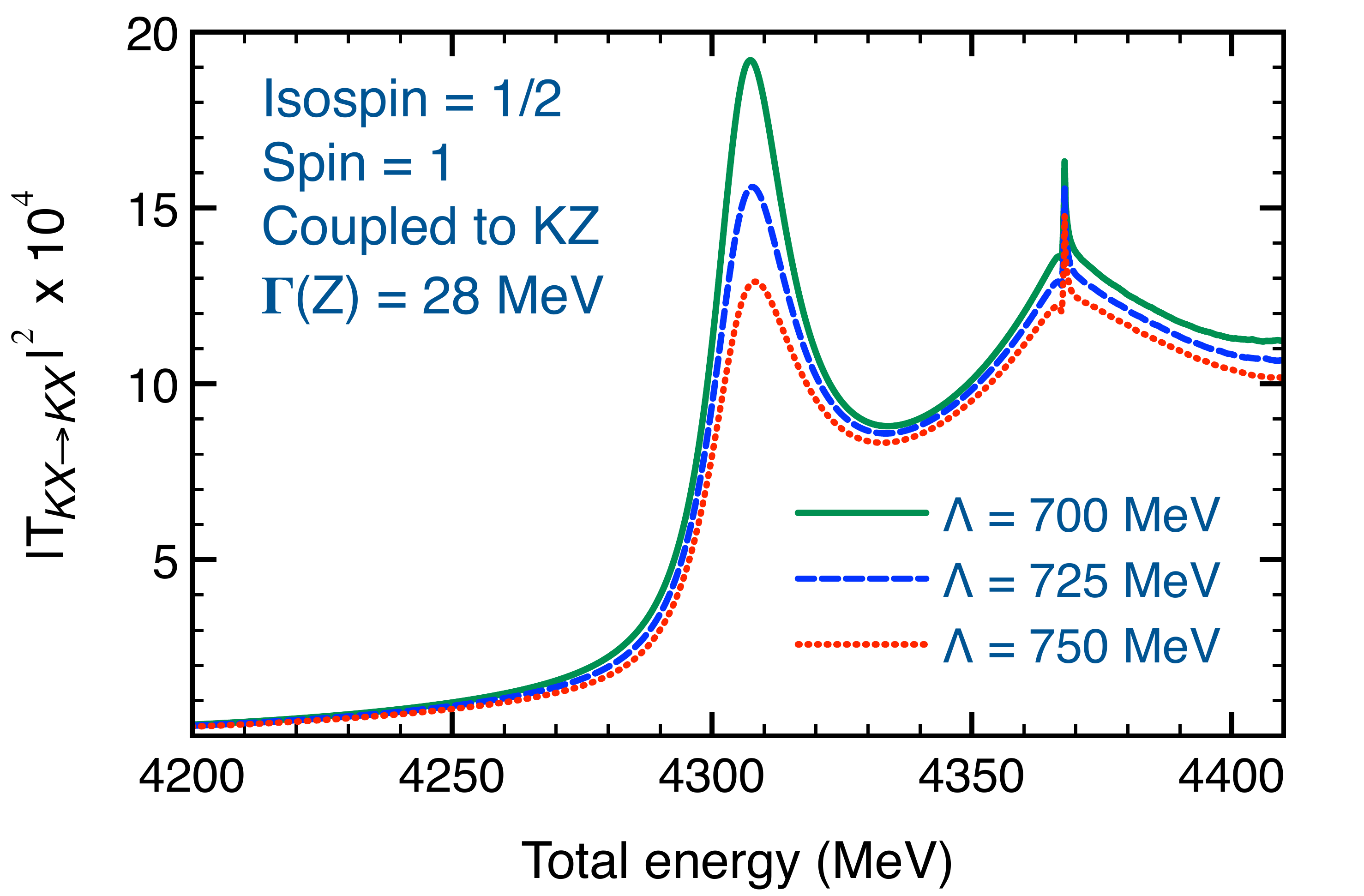}  \includegraphics[width=0.47\textwidth]{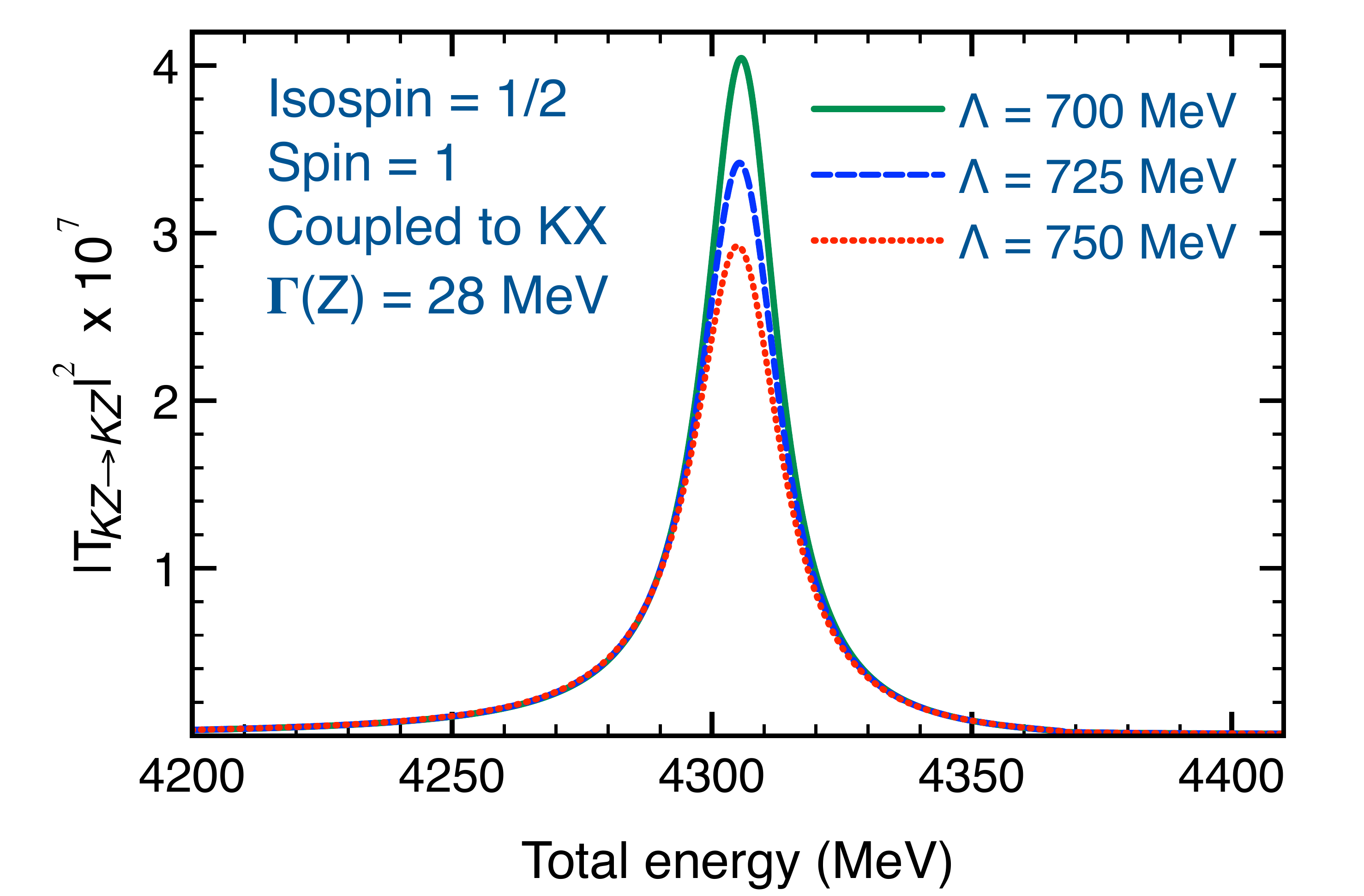}
\caption{Modulus squared of the $KX$ and $KZ$ scattering amplitudes in $I=1/2$. These results have been obtained by solving scattering equations while treating $KX$ and $KZ$ as coupled channels and considering $\Gamma(Z)=28$ MeV. A cusp related to the three-body $KD\bar D^*$ threshold is observed in the $KX\to KX$ amplitude.}
\label{Fig:KXKZ}
\end{center}
\end{figure}

We have studied the effect of coupling the $KX$ and $KZ$ systems and, thus, allowing the transitions between them. The modulus squared amplitudes obtained for both systems, by solving the scattering equations within a  coupled channel approach, are shown in Fig.~\ref{Fig:KXKZno} for the case in which the width of $Z$ is neglected (see the footnote~\ref{foot} for the origin of the width found for the peaks) and in Fig.~\ref{Fig:KXKZ} considering $\Gamma_Z=28$ MeV. As can be seen from these two figures, the consideration of the nonzero width of the $Z$ state has an impact on the results, especially on the $KX$ amplitude, even if the peak position remains almost unaltered.

Taking into account the width of $Z$ when coupling $KX$ and $KZ$, the mass and width of the state found in the $KX$ configuration can now be written as $M - i \Gamma/2 = (4308 \pm 1) - i (8 \pm 1)$ MeV, and of the $KZ$ configuration is $M - i \Gamma/2 = (4306 \pm 1) - i(9 \pm 1)$ MeV. We also find that, at the peak position, the magnitude of the squared amplitude obtained when the three-body system is rearranged as $KZ$ is around 200 times bigger than that found when the system rearranges itself as $KX$. Notice that the coupled channel scattering has shifted the peak positions in the uncoupled $KX$ and $KZ$ amplitudes such that now a peak is obtained, basically, at the same energy in both cases. Our findings, thus, imply that a $K^*$ meson around 4307 MeV should be observed in experimental investigations.

In summary, we have studied the $KD\bar D^*$ systems where the $D \bar D^*$ is treated as a cluster forming $X(3872)$ or $Z_c(3900)$.~We find that this dynamics leads to the generation of a new state of molecular nature (see Fig.~\ref{Fig:nature}) which corresponds to a $K^*$ meson with hidden charm and important $K$-$X$ and $K$-$Z_c$ components in its wave function.  The mass of the state is $(4307\pm 2)$ MeV with a width of $(9\pm 2)$ MeV.  Interestingly, a recent study~\cite{Ma:2017ery} solving the Schr\"{o}dinger equation for the $D\bar D^*K$ system, but with a very different dynamics than the one used here, found a state with a mass of 4317 MeV.

So far there is no experimental data available on $K^*$ states in the energy region investigated in the present work~\cite{Patrignani:2016xqp}, so the result found here is a prediction for a $K^*$ meson with hidden charm and of molecular three-body nature. Such state can be found at facilities, such as BEPC, in processes with final states, such as $\bar K^0 D_s^+ D^-$. We hope that our work encourages such experimental investigations.\\

The authors are grateful to Prof. Eulogio Oset for reading the manuscript and for giving very useful suggestions. This work was partly supported by DFG and NSFC through funds provided to the Sino-German CRC 110 ``Symmetries and the Emergence of Structure in QCD'' (Grant No. TRR110), the National Natural Science Foundation of China (NSFC) under Grants No. 11735003, No. 11522539, 11375024, and No. 11775099,  the Fundamental Research Funds for the Central Universities, 
 FAPESP (Grant No. 2012/50984-4), and CNPq (Grant Nos. 310759/2016-1 and 311524/2016-8).  
K.P.K and A.M.T thank Beihang University for the hospitality during their stay when this work was initiated.
\bibliographystyle{elsarticle-num.bst}
\bibliography{refs}

\begin{thebibliography}{10}
\bibitem{Zweig:1981pd}
G.~Zweig, {An SU(3) model for strong interaction symmetry and its breaking.
  Version 1}.
\end{thebibliography}

\end{document}